\documentstyle[sprocl]{article}
\def\Journal#1#2#3#4{{#1} {\bf #2}, #3 (#4)}

\def\NPB{{\em Nucl. Phys.} B}
\def\PLB{{\em Phys. Lett.}  B}
\def\PRL{\em Phys. Rev. Lett.}
\def\PRD{{\em Phys. Rev.} D}

\def\be{\begin{equation}}
\def\ee{\end{equation}}
\def\bea{\begin{eqnarray}}
\def\eea{\end{eqnarray}}

\begin{document}
\title{Lepton flavor violation and electric dipole moments from an intermediate
scale in SUSY grand unification}
\author{ N. G. Deshpande, B. Dutta, 
$^{\dagger} $  }
\address{ Institute of Theoretical Science, University of Oregon, Eugene, OR
97403  }
\author{E. Keith}
\address{Department of Physics, University of California, Riverside, CA 92521}
\maketitle\abstracts{We show how an intermediate gauge symmetry breaking scale
can be a source of lepton flavor violation and EDM of electron and neutron in
SUSY GUTs. The Universal soft supersymmetry breaking operators can be
introduced at the GUT scale or above.\\{\bf[OITS-614,UCRHEP-T178]}}
Supersymmetric grand unified theories (GUTs) with intermediate gauge symmetry
breaking scales are attractive for many reasons. For example, in models
where the intermediate breaking scale
$M_I \sim 10^{10}-10^{12}$ GeV, one can naturally  get a neutrino mass in the
interesting range of
$\sim 3-10$ eV, also this window is of the right size for a  PQ-symmetry to
be broken so as to solve the strong CP problem. Models which allow even lower
intermediate gauge symmetry breaking scales
\cite{[DKR],[EMa]} e.g.
$M_I\sim 1$ TeV are also interesting since they predict relatively light new
gauge fields. 

In this talk, mainly based on  Ref. \cite{[1],[2],[3]}, we will discuss some
additional features of the intermediate scale e.g. lepton flavor violation and
the production of an electron and neutron electrec dipole moment (EDM). We will
always assume that supersymmetry is broken via soft breaking terms introduced at
a super high scale and these terms are flavor blind  and  CP invariant. As
examples, in Ref.
\cite{[1],[2],[3]} we have used four different models with intermediate gauge
symmetry breaking scales Ref.\cite{[DKR],[EMa],[ML]}. 

It has recently been pointed out \cite{[LJdhAS]} that significant lepton
and quarkflavor violation, as well a electron and neutron EDMs, can arise in
supersymmetric (SUSY) grand unified theories. The origin of this flavor
violation resides in the largeness of the top Yukawa coupling and the assumption
that supersymmetry is broken by flavor uniform soft breaking terms communicated
to the visible sector by gravity at a scale $M_X$. Assuming that $M_X$ is the
reduced Planck scale which is much greater than  $M_G$, renormalization effects
cause the third generation multiplet of squarks and sleptons which belong to the
same multiplet as the top in the grand unified theory (GUT) to become lighter
than those of the first two generations. The slepton and the charged lepton mass
matrices can no longer be simultaneously diagonalized thus inducing lepton
flavor violation through a suppression of the GIM mechanism in the slepton
sector.  The evolution of soft terms from $M_X$ to $M_G$ causes these flavor
violations, which disappear when $M_X=M_G$. Here, we explore another class of
theories which are SUSY SO(10) GUTs which break down to an intermediate gauge
group $G_I$ before being broken to the Standard Model (SM) gauge group at the
scale $M_I$. In this class of theories, even if
$M_X=M_G$, lepton flavor violation arises due to the effect of the third
generation neutrino Yukawa coupling on the evolution of the soft leptonic terms
from the grand unification scale to the intermediate scale. Depending on the
location of the intermediate scale
$M_I$ and the size of the top Yukawa coupling at $M_G$, these rates can be
within one order of magnitude of the current experimental limit. Our results
 also indicate that if
$M_X>M_G$ in SUSY SO(10) models with an intermediate scale, the predicted rates
of lepton violating processes are further enhanced. We will concentrate on the
decay
$\mu\rightarrow e\gamma$ as an example since experimentally it is likely to be
the most viable.

With $G_I$=SU(2)$_L\times$SU(2)$_R\times$SU(4)$_C$ ($\{ 2_L\, 2_R\, 4_C\}$), the
quarks and leptons are unified. Hence, the $\tau$-neutrino Yukawa coupling is
the same as the top Yukawa coupling. Through the renormalization group equations
(RGEs), the effect of the large
$\tau$-neutrino Yukawa coupling is to make the third generation sleptons lighter
than the first two generations, thus mitigating the GIM cancellation in one-loop
leptonic flavor changing processes involving virtual sleptons. Although the
quarks and leptons are not unified beneath the GUT scale when
$G_I$=SU(2)$_L\times$SU(2)$_R\times$U(1)$_{B-L}\times$SU(3)$_c$ ($\{ 2_L\, 2_R\,
1_{B-L}\, 3_c\}$), the same effect is produced from the assumption that the top
quark
 Yukawa coupling is equal to the $\tau$-neutrino Yukawa coupling at the GUT
scale. In Ref. \cite{[1],[2]}, one can see that such models can easily predict
rates of lepton flavor violation that are within an order of magnitude beneath
experimtal limits and can sometimes even put limits on the allowable parameter
space. 

When we calculate the EDM of the electron or neutron we consider the phases at
the  gaugino-slepton-lepton or gaugino-squark-quark vertices. In fact, whenever
there is  an intermediate scale, irrespective of $G_I$   such phases  are
generated. The reason for this is that right-handed quarks or leptons are unified
in a multiplet in a given generation. The superpotential for an intermediate
gauge symmetry breaking model can be written (when $G_I=\{ 2_L\, 2_R\, 4_C\}$)
as $W_Y={\bf \lambda_{F_u}}{ F}{ \Phi_2}{ {\bar F}} +
 {\bf\lambda_{F_d}}{ F}{ \Phi_1}{{\bar F}}$,  where $F$ and ${\bar F}$ are the
superfields containing the standard model fermion fields and transform as
$(2,1,4)$ and $(1,2,{\bar 4})$ respectively and we have suppressed the
generation and gauge group indices. We
choose to work in a basis where
${\bf
\lambda_{F_u}}$ is diagonal in which $W_Y$ can be expressed as $W_Y={ F}{\bf
{\bar\lambda_{F_u}}}{ {\bar F}}{
\Phi_2} +
 { F}{\bf {\bf U^*}{\bar\lambda_{F_d}}}{\bf U^{\dag}}{{\bar F}}{ \Phi_1}$. The matrix {\bf U}
is a general
$3\times3$ unitary matrix with 3 angles and 6 phases. It can be written as 
${\bf U}={\bf S^{\prime *}} {\bf V} {\bf S}$,  where {\bf V} is the CKM matrix and ${\bf S}$
and
${\bf S^\prime}$ are  diagonal phase matrices. At the weak scale,  
 we have the Yukawa terms:
\begin{eqnarray} W_{\rm MSSM}&=&Q{\bf \bar\lambda_u}{\bf U^c}H_2 +Q{\bf V^*}{\bf
\bar\lambda_d}{\bf S}^2{\bf V^{\dag}}D^cH_1\nonumber\\ &+&E^c{\bf V_I^*}{\bf
\bar\lambda_L}{\bf S}^2{\bf V_I^{\dag}}LH_1\, ,
 \end{eqnarray} where 
 ${\bf S}^2\equiv {\rm Diag.}\left(
               {\rm e}^{i\phi_d}, {\rm e}^{i\phi_s}, 1   \right)$ 
 is a diagonal phase matrix with two independent phases, and ${\bf V_I}$ is the
CKM matrix at the scale $M_I$.  When
$G_I=\{ 2_L\, 2_R\, 1_{B-L}\, 3_c\}$,  the additional CKM-like phases will be
generated in exactly the same way  as described above. Results for specific
examples can be found in Ref. \cite{[3]}, where one can see that highly
significant EDMs can be produced by the physics of the intermediate scale  gauge
symmetry. How the b-$\tau$ unification hypothesis works for these models has
been discussed in the Ref. \cite{[3]}.

Another interesting feature of these models is that they contribute to the
$B-{\bar B}$ and
$K-{\bar K}$ mixing with a phase structure which is different from the standard
model
\cite{[DDO]}. Consequently the prescribed modes for measuring CKM would measure
something different. For example, the $B_d$ decay modes such as
$\pi \pi$, $\psi K_S$ and $D^0 K^*$(892) which are expected to be preferable in
experiment and would extract $\alpha$, $\beta$  and $\gamma$, respectively, in
the SM, however with intermediate scale, these modes would measure
$\pi-(\beta +\gamma +\phi_{B_d}/2)$, $\beta +(\phi_{B_d} +\phi_K)/2$ and
$\gamma$, respectively.   However a complete study of $B_d$ and $B_s$ decay
modes as presented in the ref.\cite{[DDO]} can extract the CKM as well as the
new phases.


\begin{thebibliography}{99}

\bibitem{[DKR]}N. G. Deshpande, E. Keith and T. G. Rizzo, \Journal {\PRL}{70}{
3189}{1993}; E. Ma, \Journal {\PRD}{51}{236}{1995}.

\bibitem{[1]}N.G. Deshpande, B. Dutta, and E. Keith, \Journal {\PRD}{54}{R730}
{1996}.

\bibitem{[2]}N.G.Deshpande, B. Dutta, and E. Keith,hep-ph/9604236,(\Journal
{\PLB}{}{}{1996} in Press). 

\bibitem{[3]}N.G. Deshpande, B. Dutta, and E. Keith,  hep-ph/9605386, (\Journal
{\PLB}{}{}{1996}in Press).
\bibitem{[ML]}D-G. Lee and R. N. Mohapatra, \Journal
{\PRD}{52}{4125}{1995}.  

\bibitem{[LJdhAS]} R. Barbieri and L. J. Hall, \Journal {\PLB}{338}{212}{1994};
S. Dimopoulos and L. J. Hall, \Journal {\PLB}{344}{185}{1995};  R. Barbieri, L.
J. Hall, and A. Strumia, \Journal {\NPB}{445}{219}{1995}; B. Dutta and E. Keith,
\Journal {\PRD}{52}{6336}{1995}.

\bibitem{[DDO]} N.G. Deshpande, B. Dutta, and S. Oh,
hep-ph/9608231, (\Journal
{\PRL}{}{}{1996} in Press).  

\end{thebibliography}
\end{document}